\documentclass[aps,pre,twocolumn,floatfix]{revtex4}
\usepackage{graphicx,color}
\usepackage{amsmath,amssymb}

\newcommand{\figrepl}{
\begin{figure}[htbp]
        \centering
        \includegraphics[width=3.25in]{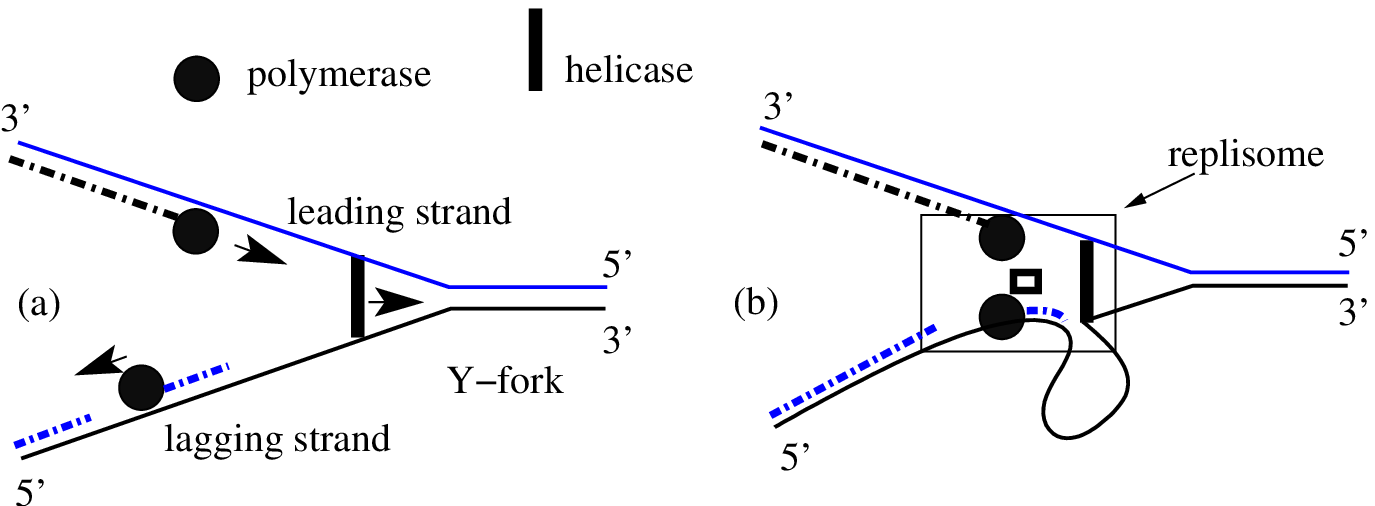}
        \caption{ Schematic diagram of the replication process. Only
          the helicase and the polymerases are shown with arrows
          indicating directions of motion. The solid lines are the
          parent DNA strands with 3', 5' ends marked.   The
          dash-dot lines represent the newly synthesized DNA. In (a),
          the members of the replication machinery move on the DNA
          individually. The small fragments on the lagging strand are
          the Okazaki fragments.  The replisome model is shown in (b).
          The small thick lined box is the clamp loader that loads the
          lagging strand DNA on the replisome.  In both models, the
          whole machinery works near the Y-fork. }\label{fig:repl}
\end{figure}
}

\newcommand{\figrecg}{
\begin{figure}[htbp]
        \centering
        \includegraphics[width=3.25in]{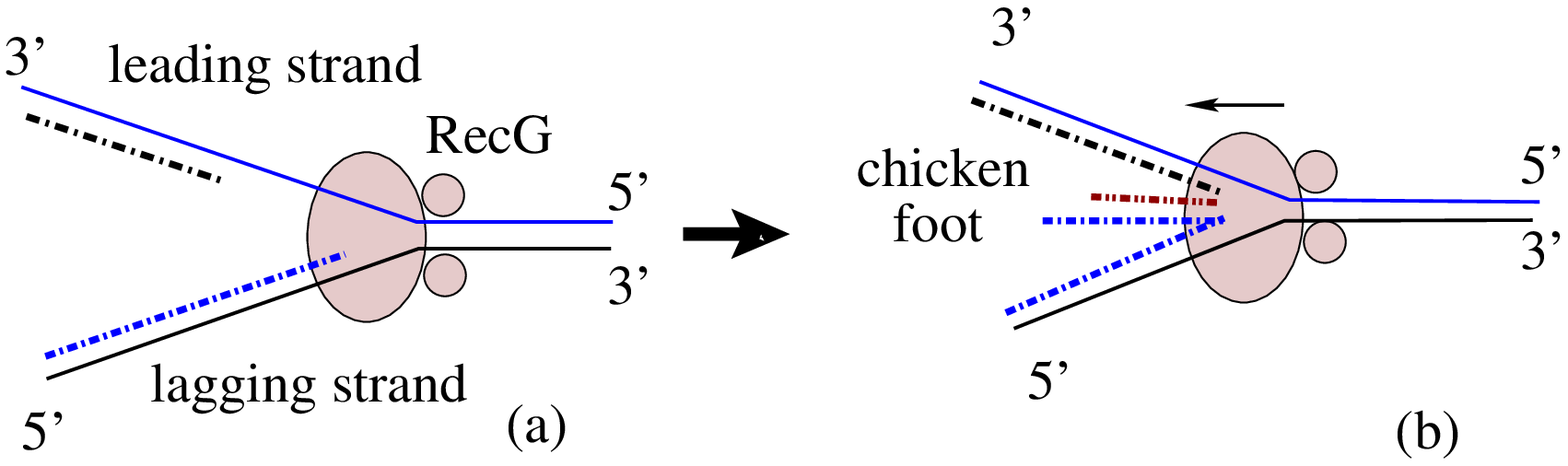}
        \caption{ Schematic diagram of the repair process by RecG
          (shaded ellipse and two circles) at the Y-fork. (a) RecG at
          the stalled Y-fork with a longer lagging strand pair.  (b)
          Fork reversal and chicken-foot configuration (Holliday
          junction). The arrow denotes the direction of motion of the
          fork.  The four arm junction is created by the fork reversal
          and the continuation of the leading strand pair by making use of
          the pair on the lagging strand.  }\label{fig:recg}
\end{figure}
}

\newcommand{\figfree}{
\begin{figure}[htbp]
        \centering
        \includegraphics[width=1.35in]{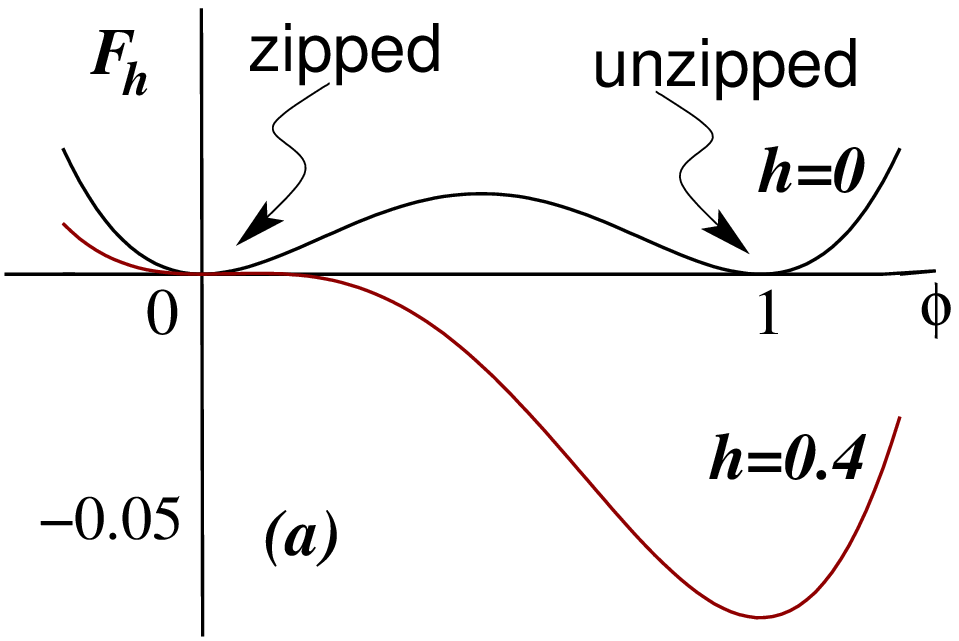}
       \hfill \includegraphics[width=1.35in]{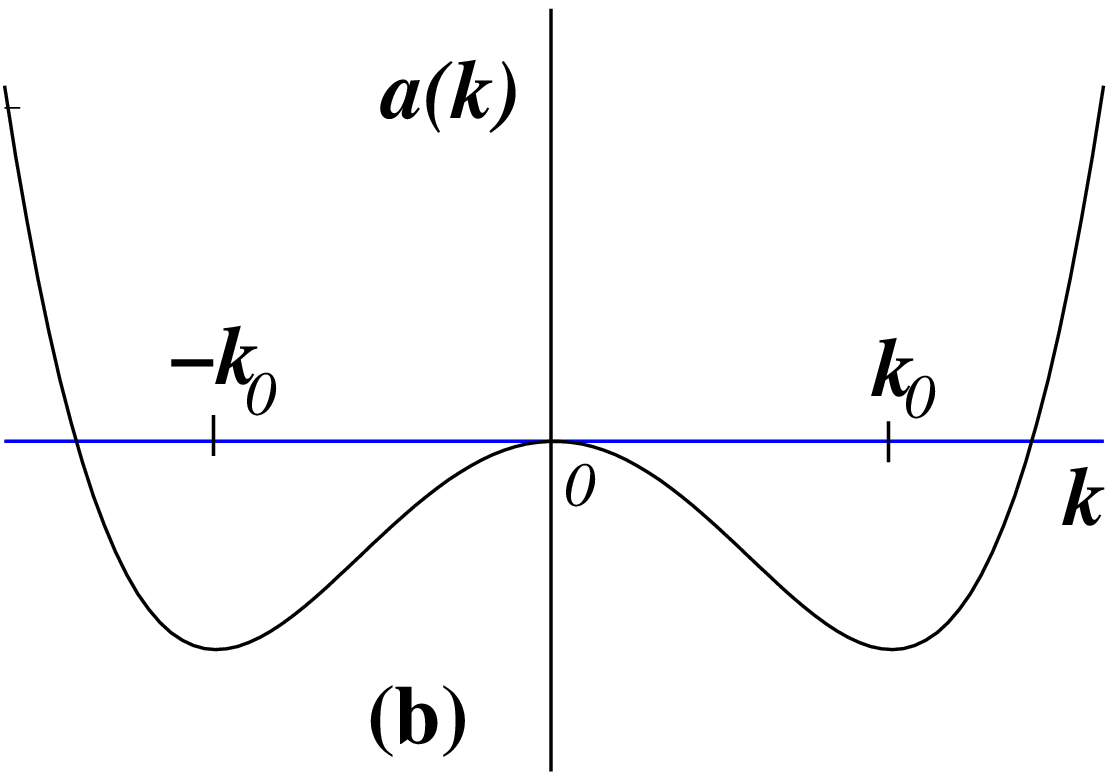}
       \caption{ (a) $F_h(\phi)$ vs $\phi$ for $h=0$ and $h=0.4$.  The
         perturbation by $h$ makes the unbound state the stable phase.
         (b) $a(k)$ vs $k$ for the $D'<0$ case. The minima at $k=\pm
         k_0$ suggest that the stable state is not the homogeneous
         state with $\phi=0$ or $1$, but a different  state with a
         modulating wave-vector $k=k_0$.  }\label{fig:free}
\end{figure}
}

\newcommand{\figprof}{
\begin{figure}[htbp]
        \centering
        \includegraphics[width=3.25in]{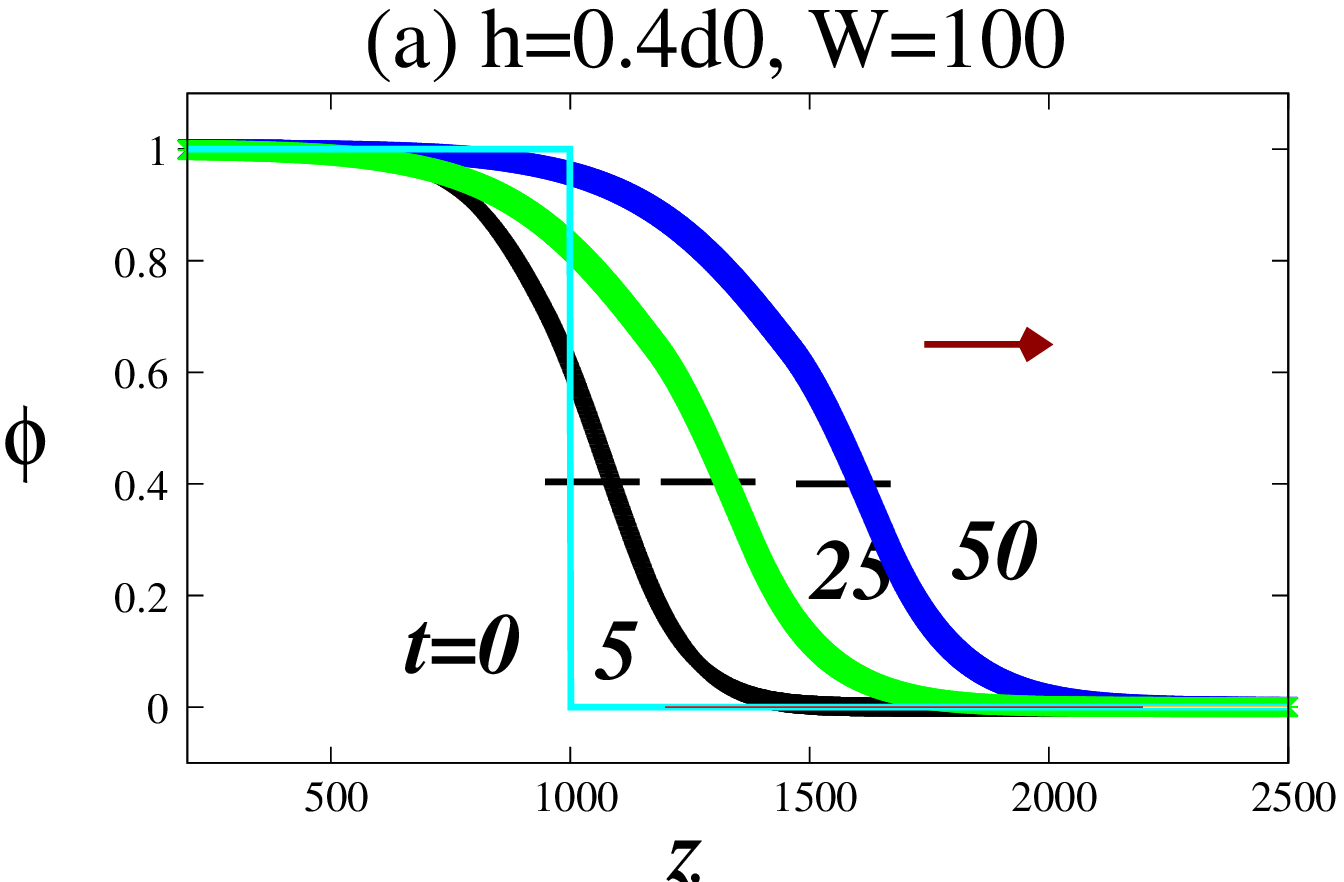}\\[18pt]
        \includegraphics[width=3.25in]{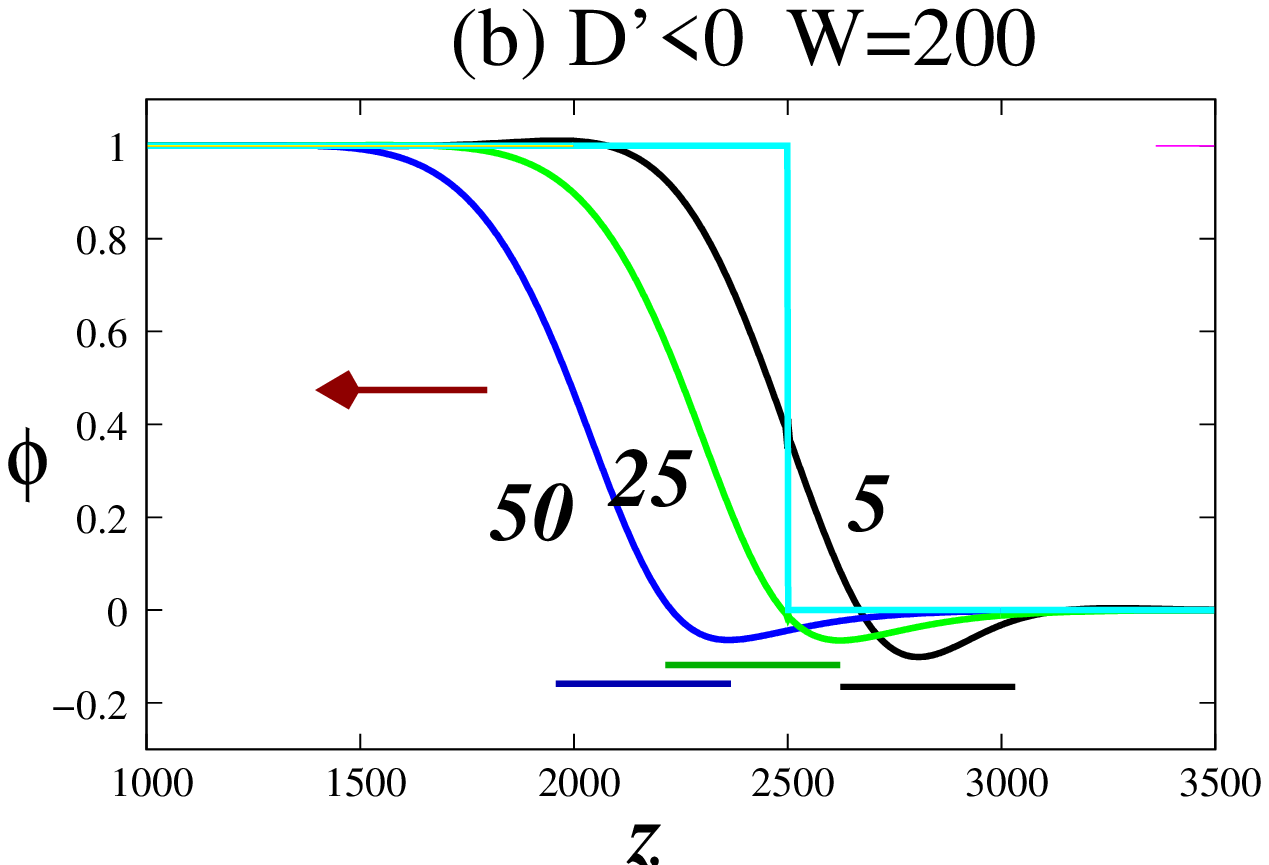}
        \caption{ $\phi$ vs. $z$ profile for various times as marked.         
          Both $z$ and $t$ are discretized. (a) For general helicase,
          Eq.  \ref{eq:3} and (b) for RecG-type case for Eq.
          \ref{eq:12}.  The horizontal bar indicates the location of
          the perturbation.  In (a) the helicase bites at $\phi=0.5$
          while in (b) the bite is at $\phi=0.1$, in both cases over a
          length $2W$.  We see the velocity (direction indicated by an
          arrow) towards the zipped side in (a) but in the opposite
          direction in (b).  The data are for $D=\gamma=1, h_R=2$, so
          that $D'<0$.
        }\label{fig:prof}
\end{figure}
}

\newcommand{\figvel}{
\begin{figure}[htbp]
        \centering
        \includegraphics[width=3.25in]{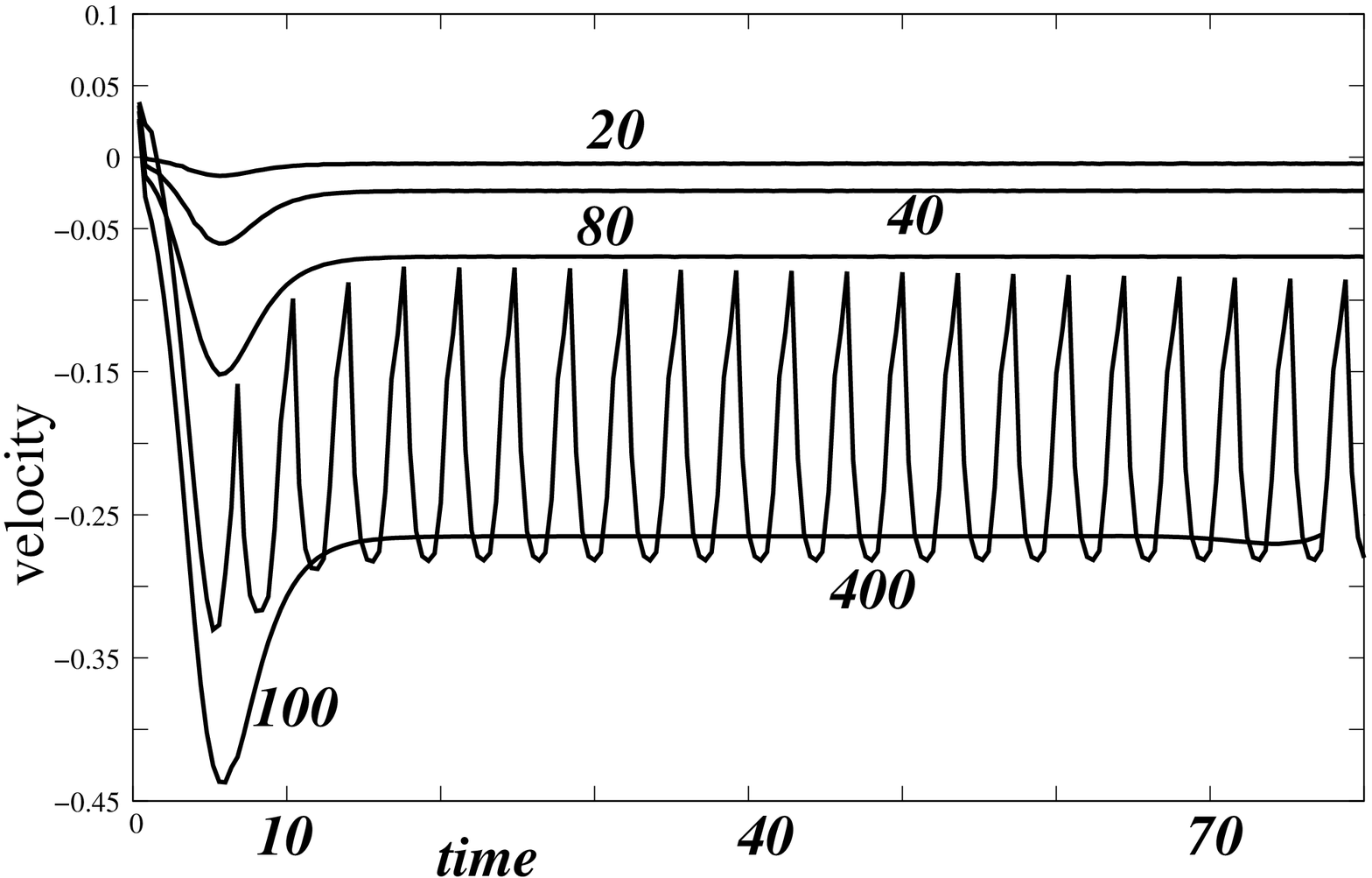}
        \caption{Velocity as a function of time for the RecG type
          case. because of the continuous change of the profile, the
          velocity may show a time dependence as seen in the case marked
          400 (the spatial width of the perturbation in arbitrary
          units). Same situation as in Fig. \ref{fig:prof}b.  }\label{fig:vel}

\end{figure}
}

\begin{document}

\title{Interfacial instability  and  DNA fork reversal by repair proteins}
\author{Somendra M. Bhattacharjee}
\email{somen@iopb.res.in}
\affiliation{Institute of Physics, Bhubaneswar 751005 India.}
\date{\today}
\begin{abstract}
  A repair protein like RecG moves the stalled replication fork in the
  direction from the zipped to the unzipped state of DNA.  It is
  proposed here that a softening of the zipped-unzipped interface at
  the fork results in the front propagating towards the unzipped side.
  In this scenario, an ordinary helicase destabilizes the zipped state
  locally near the interface and the fork propagates towards the
  zipped side.  The softening of the interface can be produced by the
  aromatic interaction, predicted from crystal structure, between RecG
  and the nascent broken base pairs at the Y-fork.  A numerical
  analysis of the model also reveals the possibility of a stop and go
  type motion.
\end{abstract}


\maketitle

\section{Introduction}
The semi-conservative replication of DNA requires its unzipping by
helicases, and synthesis of new strands over the opened parent strands
by dna-polymerases (DNA-polymerase~III) with high
processivity\cite{lehn,baker98}.  The basic features of replication
are as follows. The two strands of DNA run in opposite directions.
The polymerase with unidirectional 3'-5' motion can then follow the
helicase on one strand (the leading strand) only, but not on the other
(the lagging strand).  Though the two polymerases are identical, the
lagging strand polymerase performs a more complicated job, e.g. (i) it
makes the new strand in pieces (Okazaki fragments), (ii) it has to
shift repeatedly to newer positions closer to the moving helicase,
(iii) it restarts the polymerization process anew after every shift,
and, then (iv) the small Okazaki fragments are to be joined by dna
ligase.  In addition primase is required for initiation of the Okazaki
fragments, sliding clamp proteins to tether the polymerases to DNA, to
name a few more.  Moreover, the fidelity of replication requires
additional repair or proof-reading capabilities which need to become
functional as and when needed.  Processivity in this context is
defined as the number of base pairs added at a stretch each time the
replication machinery binds to DNA (for E.  Coli, $4.6\times 10^6$
base pair long DNA is ultimately replicated in 40mins).  It is
understood that most of the polymeric molecules of the replication
machinery can work independently in vitro but all of these need to
work in tandem over a long time and long distance (along the DNA) for
the processivity observed during replication.

The controversy with the replication dynamics can be summarized in the
following way\cite{lehn,baker98}. In one class of models, the
replication machineries move on the DNA. See Fig. \ref{fig:repl}a.
One needs tight coupling between the lagging strand polymerase and the
helicase, though they are independent, to facilitate repeated
recognition of the newly unzipped region of DNA.  Not only the
polymerase and helicase, others like the primase, ligase also need to
get correlated on the lagging strand.  The antithesis is the proposal
of a replisome - a complex of all the objects staying together with
the DNA looping through the complex in a very particular
manner\cite{odonnell}. See Fig. \ref{fig:repl}b.  Since the DNA goes
through the replisome, there is expected to be a depletion layer of
nutrients surrounding a replisome and it needs to be supplemented by a
current towards it.  More perplexing is that, if a complex can exist
for a long enough time during replication, then why it is so elusive
in vitro and what is responsible for the ``bound state'' in vivo.

\figrepl

Single molecular experiments have revealed correlations between
synthesis by T7 DNA polymerase and T7 helicase activity on a dsDNA,
e.g. efficient duplex DNA synthesis require combined action of
polymerase and helicase but no direct or specific interaction between
the polymerase and helicase play any role\cite{patel}. For
bacteriophage T4, the assembly of the polymerase and the clamp-loader
with DNA has been found to be highly dynamic and well
coordinated\cite{benko}.  The ``replisome'' may itself be dynamic in
nature with not just two but even three polymerases, and others like
repair factors can be dynamically attached to it\cite{lovett}.  In
case the replication process were tightly controlled, hard-coded in
the functioning of the molecules, then that would be reflected in the
statistics of the lengths of the Okazaki fragments on the lagging
strand as well.  The probability distribution of the lengths of these
fragments for T4 and T7 have been found to be very broad and highly
non-gaussian, compared to very sharp distributions expected from
individual primase controlled models\cite{okazaki}.  Purely stochastic
models have been proposed to study correlations in Okazaki fragment
distributions\cite{cowan}.  These newer experimental methods and
analysis, apart from elucidating the nature of the complexity of
replication, are also bringing out the importance of dynamics and
correlations (also called coordination) of the protein complexes and
DNA strands in particular near the fork.

\figrecg

An important aspect of the replication process is the repair mechanism
also involving helicases.  A helicase is a motor protein that
generally facilitates in opening the DNA and leads the replication
machineries.  Such a helicase moves from the unzipped (i.e., the open
side) to the closed side.  There are actually many helicases, e.g.
almost 20 in E. Coli (though not clear why so many), but they
generically are known to have a part (called the wedge domain) that
maintains the two strands of the DNA at a distance much larger than
the base pair distance.  The motor action (e.g. for dnaB\cite{dnab})
and often additional pulling action (e.g.  PcrA\cite{pcra}) carry the
opening process on.  It stops in case there is a nick or a lesion on
the leading strand.  The replication process stalls, the whole
replication assembly disperses, and a repair process takes over.
Often, RecG is the helicase that starts the repair job by performing a
fork reversal\cite{atkin}.  Based on the crystal structure, it was
suggested that the wedge domain maintains the separation but in
addition there are two domains that sit on the zipped side and pushes
the DNA towards the unzipped side thereby scooping out the DNA on the
lagging strand\cite{recg}.  See Fig. \ref{fig:recg}.  This continues
until the nascent end on the leading strand is reached, forming a
chickenfoot configuration.  The job of RecG is over, the repair
process is completed by other objects via a Holliday junction and then
the replication restarts.  In this model, RecG has two jobs, zipping
the parent strands and unzipping the nascent duplex on the lagging
strand.  The activities of a pre-assembled RecG-DNA on (i) a three way
DNA (consisting of the Y-fork with a nascent duplex on the lagging
strand) and (ii) a two-way DNA (consisting of a pure Y-fork) are found
to be the same, if initial transients are ignored\cite{martinez}.

We therefore take this view
that the main job in the repair process of RecG is the zipping of the
DNA or the fork reversal.

Our purpose in this paper is to show how a coupling of the helicase
and the replication fork (Y-fork) can be used to formulate the
backward mobility of RecG-like repair proteins, in addition to the
forward motion of general helicases.  Our hypothesis is that the
Y-fork is an essential element in the whole replication process, and
it is at the heart of the correlated dynamics\cite{epl04}. This
hypothesis is based on the observation that in both the biological
models and experiments mentioned earlier, the common feature is the
proximity of a Y-fork - the junction of an unzipped and a zipped DNA,
and dynamics  plays an important role, may even be responsible for the
formation  of the replisome.

The first step to motivate this connection is to analyze the helicase
activity.  Traditionally, DNA opening is considered a melting
phenomenon where thermal noise, fluctuations in base pair
breaking-joining, and polymer configurations play important
roles\cite{lehn}.  A different mechanism is the force induced
unzipping transition at temperatures below the melting
point\cite{smb,maren,maren2,zip}.  In the unzipping scenario, a coexistence
of the two phases separated by a domain wall, at this first-order
transition, is {\it the} Y-fork.  The wedge domain of a helicase
provides the constraint to maintain the DNA in a fixed distance
ensemble, and, therefore, a Y-fork develops.  The helicase activity is
then tantamount to setting the Y-fork interface in motion.  In this
language, the replication problem can be recast as a velocity
selection problem.  The Y-fork dynamics is described by the internal
dynamics of DNA, the helicase motion is controlled by the energy
supply and its own dynamical mechanism, and similarly for other
individuals.  Thus, each has its own characteristic velocity acting
alone, and that's the velocity one measures in vitro\cite{recbcd}.
Despite this diversity, these should have the same velocity when
together during replication.  Hence, the issue of velocity selection.
Similarly, a repair process is not just a resealing of the Y-fork
under DNA dynamics but has to be closely knit with the repair process.
The specialty of RecG comes from the crystal structure which indicates
a competition of the closed rings of the wedge domain with the nascent
broken pairs of the DNA through aromatic interaction\cite{recg,atkin}.
That both the forward and the backward motions can be handled in the
same framework lends credence to the basic hypothesis.

The outline of the paper is as follows.  Using a Landau type functional
to describe the zipped-unzipped coexistence we formulate the
propagating front equation when there is a local perturbation around
the interface.  The coupling of RecG with the interface is then
introduced and the effective dynamics written down.  This strong
coupling that can lead to the elastic or diffusive term to change
sign, stabilized by a fourth derivative term, is then studied.  This
formulation is done in Sec II.  A perturbative approach is used to
calculate, via a Goldstone mode, the velocity of propagation, the
starting point being the zero velocity coexisting phase situation.This
is done in Sec III.  A numerically exact solution of a discretized
version of the model is then presented in Sec IV.  Apart from
verification of the predictions of the perturbation theory,
nonperturbative effects are also found.  Sec V is the summary and
conclusion.

\section{Model and velocity selection}

\subsection{Dynamics}
We start with the fact that a double-stranded DNA in equilibrium below
its melting point can show phase coexistence with an interface
separating the two phases, the zipped and the unzipped phases.  A
helicase (or its wedge domain mentioned in Sec I) provides the
necessary constraint of a fixed distance ensemble to generate the
coexistence. To open the DNA this interface needs to be translocated.

\figfree

For standard helicases like dnaB, the motor action just forces through
the DNA so that as the location of the constraint of the fixed
distance ensemble shifts, the bound state region becomes metastable or
unstable with respect to the unzipped state.  For helicases like PcrA,
there is an additional pull by a hand-like branch on the DNA strand
near the Y-fork.  This is also an example of an active mechanism to
make the state near the Y-fork meta- or unstable but not in the bulk.
With this in mind, let us formulate our model in the following way.
The equilibrium phase coexistence - Y-fork - by a static helicase, can
be described by a Landau like free energy $F_h(\phi)$ where $\phi$ is
a scaled distance.  By choosing $\phi=0$ as the bound state and
$\phi=1$ as the open state, with a slightly negative $\phi$
representing an overtight state, we take
\begin{equation}
  \label{eq:1}
  f_h(\phi)\equiv -\frac{dF_h(\phi)}{d\phi}= \phi (\phi-\frac{1}{2}+h)(1-\phi),
\end{equation}-
so that the effective Landau-Hamiltonian can be written as 
\begin{equation}
  \label{eq:2}
  {\cal{F}}\{\phi\} =\int dz \left[ \frac{D}{2} \left(\frac{\partial\phi}{\partial z}\right)^2 + F_h(\phi)\right], 
\end{equation}
where $D$ is an elastic constant and 
\begin{equation}
  \label{eq:17}
 F_h(\phi)= \frac{\phi^2}{4} - \frac{\phi^3}{2} + \frac{\phi^4}{4} + h
 \left( - \frac{\phi^2}{2} + \frac{\phi^3}{3}\right ).
\end{equation}
The phase coexistence occurs  at $h=0$.  The coefficients are chosen,
for simplicity, to have the extrema of $F_h$  in a symmetrical fashion.
The perturbation $h>0$ makes the unzipped state the favoured phase.  
See Fig. \ref{fig:free}a.
The dynamics is described by
\begin{equation}
  \label{eq:3}
  \frac{\partial \phi}{\partial t} = \frac{\delta{\cal{F}}}{\delta \phi}
      =  D\frac{\partial^2 \phi}{\partial  z^2} + f_h(\phi),
\end{equation}
with the boundary condition: 
\begin{equation}
   \label{eq:15}
   \phi(-\infty)=1,\  \phi(+\infty)=0,
\end{equation}
and $h(z,t)$ is non-zero only near the interface.  In Eqs. \ref{eq:1}
and \ref{eq:3}, $f_h$, represents the effective force.  The boundary
conditions ensure that we have open or unzipped strand on the left
side while a closed terminal on the right side.  The chain-length has
been taken to be infinity.  Despite the resemblance, this is not the
Fisher-Kolmogorov problem\cite{kpp,wim} because there is {\it no} bulk
instability here.

Where is the helicase in this set of equations?  For a static case,
$h=0$, and there is no propagation.  The boundary conditions impose a
kink $\phi(z)$ that goes from $\phi=1$ to $\phi=0$, but the location
$z_0$ of the kink is arbitrary\cite{chaikin}. This is a Goldstone-like
mode; the interface or the kink separating the unzipped and the zipped
phases can be placed anywhere along the chain because neither Eq.
\ref{eq:3} cares about $z_0$, nor there is any energy cost in shifting
the interface. However, a static helicase fixes the position of the
Y-fork and therefore it kills the Goldstone mode by fixing $z_0$ where
the Y-fork would be.  The local instability is taken into account by
$h(z,t)$ which parametrizes the motor action and the active mechanism
of the helicase.  In short, the helicase at time $t=0$ is replaced by
two features: (i) the time independent boundary conditions and the
location of the initial Y-fork represent the wedge domain, and (ii)
the passive or active process of the helicase is represented by the
local instability parameter $h$.

The dynamic activity of a helicase makes $z_0$ a function of time.  A
propagating mode would imply
\begin{equation}
  \label{eq:4}
\phi(z,t)\equiv\Phi(z-ct),  
\end{equation}
with $c$ the velocity of the front. The helicase must move with the
same velocity and the perturbation too, so that
\begin{equation}
  \label{eq:14}
 h(z,t)\equiv H((z-ct)/W),\  H(z)\neq 0 {\ {\rm for}\ } |z|<1.
\end{equation}  
This matching of velocity is the velocity selection mentioned earlier.
$W$ is the width of the region over which the helicase affects the
fork.  A simpler, more practical, choice would be an implicit
definition $h\neq 0$ for $|\phi-\phi_h|\leq \delta\phi_w$.

RecG is similar to other helicases so far as the existence of a wedge
domain is concerned, but its interaction is with the broken bonds at
the interface (more like a surfactant).  In other words, unlike the
ordinary helicases that unzips, it does not produce any instability of
any phase but rather interacts with the interface.  With that in mind
we introduce a new Gaussian variable $y(z,t)$ representing the
helicase and its interaction with the interface (for which
$\partial\phi/\partial z\ne
0$) as
\begin{equation}
  \label{eq:10}
 {\cal{F}}\{\phi\} =\int dz \left[ \frac{D}{2} \left(\frac{\partial
      \phi}{\partial z}\right)^2 + F_0(\phi) + 2\lambda \, y\,
\frac{\partial\phi}{\partial z}+K y^2 \right], 
\end{equation}
with $ (K>0)$, where the subscript $0$ indicates $h=0$ in Eq.
\ref{eq:1}, i.e., there is no perturbation in the coexisting part of
the free energy.

Instead of coupled dynamics, we consider the effective dynamics of the
Y-fork by integrating out the $y$ field.  Such an integration leads to
a new effective Hamiltonian of the type Eq. \ref{eq:2}, with a reduced
elastic constant $D'=D- \lambda^2/K$.  For sufficiently large
$\lambda$, $D'$ can become negative and for stability a higher order
term is added.   The effective Landau energy can then be taken as
\begin{equation}
  \label{eq:18}
 {\cal{F}}\{\phi\} =\int dz \left[\frac{D'}{2} \left(\frac{\partial
      \phi}{\partial z}\right)^2 +\frac{\gamma}{2} \left(\frac{\partial^2
      \phi}{\partial z^2}\right)^2+ F_0(\phi) \right].   
\end{equation}
The dynamics is given by
\begin{equation}
  \label{eq:12}
\frac{\partial \phi}{\partial t} = D \frac{\partial^2 \phi}{\partial
  z^2} -\gamma \frac{\partial^4 \phi}{\partial  z^4}- h_R\frac{\partial^2 \phi}{\partial  z^2}+ \phi(\phi-\frac{1}{2})(1-\phi),  
\end{equation}
where the helicase (perturbing) part is written explicitly as the
$h_R$ term with $h_R\neq 0$ in a region near $\phi=0$.  We use the
fact that RecG operates near the zipped side.  With negative $D'$,
there is a preference for a modulated structure.  To see the effect of
$D'<0$, we rewrite the gradient-dependent part of ${\cal F}\{\phi\}$ in
Fourier modes as
\begin{equation}
  \label{eq:19}
  {\cal F}(\phi)= \frac{1}{2}\int\: dk\: a(k)\: \phi_k \phi_{-k},
\end{equation}
where
$$\phi(z)=\int \frac{dk}{(2\pi)^{1/2}} e^{ikz}\:\phi_k, \ {\rm and}\ 
    a(k)=D' k^2+\gamma k^4.$$ 
This favours, as shown in Fig. \ref{fig:free}b, a modulated state with
wave-vector $k_0= \gamma/|D'|$.  Such a modulated structure would be a
state with bubbles - but that does not happen here because the
perturbation is only local, meant to destabilize the interface.

The difference between the two types of helicases is now seen in the
respective $h$-type terms.  In Eq. \ref{eq:3}, the $h$-term affects
the stability of the phase while in Eq. \ref{eq:12} it affects the
interface.

\section{Perturbative approach}
We now treat the helicase effect as a small perturbation and show the
change in the direction of velocity in the two cases in a first order
perturbation theory.   
Let us take Eq. \ref{eq:2} for illustration.
If there is no perturbation, then the
free-energy, Eqs. (\ref{eq:1},~\ref{eq:2}), suggest a static
interface, i.e., velocity  $c=0$.
The profile with small perturbation  can therefore be written as 
\begin{equation}
  \label{eq:16}
 \phi =\phi_0(z) + \delta\phi(z-ct,t) 
\end{equation}
where the zeroth order solution is the static solution satisfying
\begin{equation}
  \label{eq:5}
D \frac{\partial^2 \phi_0}{\partial  z^2} + \phi_0(\phi_0-\frac{1}{2})(1-\phi_0)=0.
\end{equation}
For the boundary conditions of Eq. \ref{eq:3}, Eq. \ref{eq:5}
gives the static kink solution $\phi_0(z)$ that gives the
$z$-dependent  profile near the interface\cite{chaikin}.
The perturbed part satisfies to first order an inhomogeneous
differential equation
\begin{equation}
  \label{eq:6}
   \left[\frac{\partial \ }{\partial t} - D \frac{\partial^2 \
  }{\partial  z^2} - f'(\phi_0)\right ] \delta\phi= \delta
f(\phi_0)=h\phi_0(1-\phi_0)
\end{equation}
where $f'(\phi_0) = df/d\phi|_{\phi=\phi_0}$.  This solution of this
equation can be written in terms of the Green function\cite{wim} 
\begin{equation}
  \label{eq:7}
  \delta\phi=\int_{t_0}^t  d\tau \int_{-\infty}^{\infty} d\xi\: G(t,\tau,z,\xi) \   \delta f(\phi_0)
\end{equation}
Moreover, at first order, the velocity $c$ is small and the perturbed
part can be written as
$$\delta\phi=\frac{d\phi_0}{dz}\  (t-t_0) c.$$   
This shows that the velocity can be determined from the coefficient of
the term linear in $t$.

The Green function has an eigen-function expansion for the spatial
part
\begin{equation}
  \label{eq:8}
  \left[D \frac{\partial^2 \ }{\partial z^2} + f'(\phi_0)\right]\psi_n=E_n\psi_n,
\end{equation}
where $E_n$ are the eigen-values and $\psi_n$ the corresponding
eigenfunctions.  Since we are considering a dissipative system, it is
guaranteed that $E_n\ge 0$.  For all the eigen-functions with $E_n>0$,
the Green function will have a time contribution $G(t,t_1,z,z_1)\sim
\exp(-E_n(t-t_1))$ so that in the long time limit, these modes will
not contribute, except for the initial transients.

The Goldstone mode corresponds to a solution with $E_0=0$ as can be
verified directly with $\psi_0=d\phi_0/dz$.  This zero mode produces
the linear term in the solution (not an exponential decay in time) and
the velocity of the front comes from this term only.  The velocity
then is given by 
\begin{equation}
  \label{eq:9}
c=\frac{\int \phi_0'(z_1) (\delta f(z_1)) dz_1}{\int [\phi_0'(z_1)]^2
  dz_1}.
\end{equation}
If the helicase perturbation is operational in a region between
$\phi=\phi_+$ to $\phi=\phi_{-}$, then the velocity can be written as
 \begin{equation}
   \label{eq:11}
   c= D \frac{\Delta F}{{\cal E}},
 \end{equation}
where ${\cal E} =\int [\phi_0'(z)]^2  dz$ is the kink energy and 
$$\Delta F=\int_{\phi_{-}}^{\phi_{+}}\delta f(\phi) d\phi$$
 is the free energy
cost per unit length in the region of the bite.  The expression for
the kink energy follows from
Eqs. (\ref{eq:5}) and (\ref{eq:10}).
The velocity  found in Eq. \ref{eq:11} is
positive, as it should be.

This procedure can be implemented for the RecG case, Eq. \ref{eq:12}.
The instability of the interface now leads to a moving front but with
a negative velocity at least in the perturbative regime.  The
perturbation theory via the Goldstone mode yields
\begin{equation}
  \label{eq:13}
 c= -h_R \frac{1}{{\cal E}}\left( \frac{\partial \phi_0}{\partial  z}\right)^2\Bigg
|_{-}^{+},  
\end{equation}
where the limits are the two end points of the bite.  What we see is a
negative velocity because of the reduction of the elastic constant and
the curvature of the profile of the zeroth order interface near
$\phi=0$.  By symmetry one would get a positive velocity if the
perturbation were at $\phi=1$.

\figprof

\subsection{Numerical Solution}
The perturbation theory is around the static solution and need not be
valid in practical situations.  More structures can be expected if we
look at the nonperturbative effects of the helicase term in presence
of the dynamically generated front.  To do so, we solved the
discretized versions of the equations numerically via an implicit
procedure. The arbitrary lattice spacing $\Delta z$ and time 
$\Delta t$ are chosen small for convergence.  We have taken $\Delta
t=10^{-4}$ and $\Delta z=10^{-2}$.   All the other parameters are chosen in
these units.

The results are shown in Fig.  \ref{fig:prof} and \ref{fig:vel} in
discretized $z$ and time $t$.  The location $z_0$ of the interface is
taken to be the point where $\phi=0.5$.  We start with a sharp
interface at time $t=0$, with $z_0$ chosen away from the boundary.
Initial transients allow the interface to evolve to its natural shape.
This shape in the steady state has been found not to depend on the
initial profile.  The perturbations are applied over a region as
indicated in Fig.  \ref{fig:prof}.  For Fig.  \ref{fig:prof}a, the
local perturbation favours the unzipped state as shown in Fig.
\ref{fig:free}a while in Fig.  \ref{fig:prof}b, $h_R$ perturbation
makes both states unstable (Fig \ref{fig:free}b).  The velocity
measurements were done once the steady state is reached and the
interface is not close to the boundaries to avoid finite size or
boundary effects.

A check on the perturbation theory would be the linearity of the
velocity for small perturbations.  We have verified this and also the
fact that for a wide perturbation (much larger than the width of the
interface) the velocity should saturate to the bulk value\cite{epl04}.
These verifications are not shown here.  Fig.  \ref{fig:prof} shows
the change in the direction of the velocity by the two perturbations.
In (a) the helicase bites at a particular value of $\phi=0.5$, but
since the perturbation makes the $\phi=0$ state metastable, the
propagation is always towards the zipped side - opening of DNA.  In
case (b), the bite is also at a particular value of $\phi$ but close
to the bound state side.  We chose $\phi=0.1$.  Owing to the
instability of the profile, there is a push towards the unzipped side.
The profile in Fig. \ref{fig:prof}b shows the structure developed -
the monotonicity of the profile of Fig.  \ref{fig:prof}a is gone.
This non-monotonicity can lead to new phenomenon not captured by the
firstorder perturbation. Depending on the width of the bite, there may
even be a cancellation in Eq.  \ref{eq:13} as we see in the case
marked 400 (ignoring the initial transients) In this situation, the
helicase would have a stop-and-go type motion, of course towards the
unzipped side.  Though RecG may not do so, but there are other
helicases like RecQ, RecA which are known to have such peculiar
motions.  Therefore the proposed destabilizing mechanism can lead to
varieties of motions.  Moreover, these parameters provide us with a
set of values to classify or differentiate the helicases
quantitatively.

\figvel

\section{Summary}
A few comments can be made here. In the case of front propagation due
to bulk instability, almost a century old problem\cite{kpp}, there is
a pushed to pulled type dynamic transition as reflected in the nature
of the velocity of propagation\cite{wim}.  In our approach, the
biochemically controllable width $W$ within which a helicase works
actually acts as a finite length cut-off for the bulk transition, and,
therefore, it provides a new testing ground for any finite size
scaling of the bulk nonequilibrium transition. This would be an
extremely interesting situation where a biological problem can shed new
lights on an age-old physics/material science problem.  

Biological models have so far ignored the importance of the interface
or treating the Y-fork as a coexistence.  We hope our results would
motivate new experiments to measure the elastic energy of the
interface ($D$), possibly via independent propagation velocity
measurement. In addition, the helicases need to be classified by the
nature of the perturbation and its strength and width.  One needs to
wait for these independent experimental results for any quantitative
comparisons between the model calculations presented here and related
biological experiments.

We summarize our results here. The Y-fork generated by a helicase, if
kept in equilibrium, is a coexistence of two phases.  Ordinary
helicases work by destabilizing the zipped phase near the interface
and the local instability leads to the Y-fork motion towards the
zipped side, thereby opening the fork.  For RecG, the aromatic
interaction between the helicase and the interface leads to a
destabilization of the interface, and this could close the DNA by a
fork motion towards the open side.  We show this by a perturbation
method that also verifies the results for the ordinary case.  A
numerically exact solution is used to study the Y-fork motion in
presence of the dynamically generated front.  In the interface
destabilization case, more complex stop-and-go type motion is found to
be possible.  Coming back to the problem of replication, our results
enable us now to use this formulation to describe the replication in a
co-moving frame with the Y-fork, bypassing a direct reference to the
helicase.  It remains to be seen if one can couple the Y-fork motion
with the polymerase activity.


\end{document}